# IO-VNBD: Inertial and Odometry Benchmark Dataset for Ground Vehicle Positioning

Uche Onyekpe[1,3*], Vasile Palade[3], Stratis Kanarachos[2] and Alicja Szkolnik[4]

[1] Institute for Future Transport and Cities, Coventry University, Coventry, United Kingdom
[2] Faculty of Engineering and Computing, Coventry University, Coventry, United Kingdom
[3] Research Center for Data Science, Coventry University, Coventry, United Kingdom
[4] Coventry University, Gulson Road, Coventry, United Kingdom
onyekpeu@uni.coventry.ac.uk, ab5839@coventry.ac.uk, ab8522@coventry.ac.uk, szkolnia@uni.coventry.ac.uk

*Abstract* – *Low-cost Inertial Navigation Sensors (INS) can be exploited for a reliable solution for tracking autonomous vehicles in the absence of GPS signals. However, position errors grow exponentially over time due to noises in the sensor measurements. The lack of a public and robust benchmark dataset has however hindered the advancement in the research, comparison and adoption of recent machine learning techniques such as deep learning techniques to learn the error in the INS for a more accurate positioning of the vehicle. In order to facilitate the benchmarking, fast development and evaluation of positioning algorithms, we therefore present the first of its kind large-scale and information-rich inertial and odometry focused public dataset called IO-VNBD (**I**nertial **O**dometry **V**ehicle **N**avigation **B**enchmark **D**ataset). The vehicle tracking dataset was recorded using a research vehicle equipped with ego-motion sensors on public roads in the United Kingdom, Nigeria, and France. The sensors include a GPS receiver, inertial navigation sensors, wheel-speed sensors amongst other sensors found in the car, as well as the inertial navigation sensors and GPS receiver in an Android smartphone sampling at 10 Hz. A diverse number of driving scenarios were captured such as traffic congestion, round-abouts, hard-braking, etc. on different road types (e.g. country roads, motorways, etc.) and with varying driving patterns. The dataset consists of a total driving time of about 40 h over 1,300 km for the vehicle extracted data and about 58 h over 4,400 km for the smartphone recorded data. We hope that this dataset will prove valuable in furthering research on the correlation between vehicle dynamics and dependable positioning estimation based on vehicle ego-motion sensors, as well as other related studies.*

## Specifications Table

| Subject | Automotive Engineering, Signal Processing, Artificial Intelligence |
|---|---|
| Specific subject area | Positioning and Tracking of Autonomous Vehicles |
| Type of data | Excel csv |
| How data were acquired | Equipment<br>• Racelogic VBOX Video HD2 CAN – Bus Data Logger (10Hz) [15]<br>• Racelogic VBOX Video HD2 GPS Antenna (10Hz)[1]<br>• Huawei P20 pro, Motorola moto G7 power and Blackberry Priv using AndroSensor Application (10Hz) [2]. |
| Data format | Raw |
| Parameters for data collection | The data was collected under a diverse number of environmental scenarios and vehicle motion states. The number of scenarios |

[*] Corresponding author Uche Onyekpe: onyekpeu@uni.coventry.ac.uk

|  | considered include bumps, hard braking, wet roads etc. See Table 4 for the full list of scenarios considered. |
|---|---|
| **Description of data collection** | The data was collected using four vehicles employing the sensors on a smartphone, GPS receiver and the sensors present in the sensor cluster of the vehicle. The smartphone data is sampled at 10 Hz with a GPS update rate of 1 Hz providing a total data size of about 2.2 million x 24, while the ECU recorded data is also sampled at 10 Hz with a total data shape of about 1.4 million x 29. |
| **Data source location** | Country: England, France, Nigeria<br>Latitude and longitude (and GPS coordinates) for collected samples/data: GPS co-ordinates are provided in the dataset. |
| **Data accessibility** | Repository name: Github.com<br>Data identification number: 2005.01701<br>Direct URL to data: https://github.com/onyekpeu/IO-VNBD |
| **Related research article** | U. Onyekpe, V. Palade, and S. Kanarachos, "*Learning to Localise Automated Vehicles in Challenging Environments using Inertial Navigation Systems (INS)*" Applied Sciences 2021, *11*(3), 1270, https://doi.org/10.3390/app11031270 |

## Value of Data

- The dataset is large-scale and diverse, and it focuses on inertial vehicle navigation under complex environmental scenarios and vehicle motion states such as varying longitudinal accelerations, hard-brakes, yaw rates, velocities, mud roads, motorways, etc. (see Table 4). The dataset consists of measurements from a rich combination of ego-motion sensors such as accelerometers, gyroscope, magnetometers, wheel encoders, force sensors, etc.
- The data is useful to research institutions and industries in the benchmarking, fast development, evaluation and testing of vehicle positioning and tracking algorithms and techniques.
- The data is useful for the robust training of supervised learning algorithms in learning the correlation between the dynamics of vehicles and their displacement, with applications in the tracking or positioning of vehicles and robots in GPS deprived environments using noisy low-cost sensors.

## Data Description

The total dataset consists of about 100 h of recorded driving data on public roads by 8 different drivers with different driving styles as defined on Table 1 , where defensive driving refers to situations where the vehicle is turned at less than 0.3 g, swerved at less than 3.3 km/hr or decelerated at less than 0.3 g, whilst aggressive driving refers to respective situations above these thresholds [3]. The data is divided into sets based on cities and towns driven via, road conditions, weather conditions, driving length and time, driving style and driving features (see Tables A1-1 to A7). The dataset also contains more than 20 min of data recorded from the stationary vehicle to aid in the estimation of the sensors' bias. To add to the diversity of the data consisting of a number of complex driving scenarios as shown on Table 4, the data was recorded with different tyre pressures. Datasets with each unique tyre pressures are indicated on Tables

A1-1 to A6 using Table 2 as a guide. Tables A1-1 to A7 reveal more detailed information on each set of the data. The data logged from the vehicle's CAN bus are denoted with the prefix "*V-*" and the smartphone data denoted with the prefix "*S-*". The "*S-*" datasets are acquired from the sensors in a smartphone attached to the vehicle mimicking its motion[*]. While all the "*V-* "datasets were collected only in England, the "*S- "datasets* were collected in England, France and Nigeria.

Over the course of the data collection, communication difficulties between the GPS receiver and satellites were encountered. Information on data indexes recorded during these periods are provided in a file titled "*GPS outages*". Where possible, the "*S-*" and "*V-*" datasets which were collected simultaneously[1], are manually synchronised and stored in the folder named "*Synchronised V and S datasets*".

Importantly, despite the effort lent towards an accurate alignment of the smartphone's sensor axis with that of the vehicle, the precision of the measurements were interfered by vehicular vibrations averagely estimated to be about 0.15 g of acceleration and 0.08 rad/s of yaw rate particularly at peculiar scenarios such as hard brakes or over bumps. Information on the amount of gravitational acceleration measured by each of the three axis are provided in the "*S-*" datasets to help in the correction of the measured acceleration. The data is stored in csv format at https://github.com/onyekpeu/IO-VNBD along with useful Python development tools.

*Table 1 Driving pattern of each driver.*

| Driver | Driving Style |
| --- | --- |
| A | Aggressive and Defensive |
| B | Aggressive |
| C | Aggressive and Defensive |
| D | Aggressive and Defensive |
| E | Aggressive and Defensive |
| F | Defensive |
| G | Defensive |
| H | Defensive |

*Table 2 Various tyre pressures experimented on.*

| Notation | Tyre Pressure (psi) |
| --- | --- |
| A | Front right - 16 |
|   | Front left - 15 |
|   | Rear right - 14 |
|   | Rear left - 14 |
| B | Front right - 31 |
|   | Front left - 31 |
|   | Rear right - 25 |
|   | Rear left - 25 |
| C | Front right - 33 |
|   | Front left - 33 |
|   | Rear right - 31 |
|   | Rear left - 27 |
| D | Front right - 33 |
|   | Front left - 33 |
|   | Rear right - 26 |
|   | Rear left - 26 |
| E | Front right – N/A |
|   | Front left - N/A |
|   | Rear right – N/A |
|   | Rear left – N/A |

---

[*] It is difficult to truly determine the centre of gravity of the car under different dynamic conditions, hence the smartphone recording approximates the true motion of the car.
[1] Not all "V-" and "S-" dataset were collected simultaneously. All the "V-" datasets without a corresponding "S-" dataset and vice-versa are not placed in the "Synchronised V and S datasets" folder.

*Table 3 Information recorded from the Ford Fiesta's ECU.*

| No | Column Heading | Unit |
|---|---|---|
| 1 | No of GPS satellites available | N/A |
| 2 | Time since start of day | seconds |
| 3 | GPS Latitude | degrees |
| 4 | GPS Longitude | degrees |
| 5 | GPS Velocity | km/hr |
| 6 | GPS Heading | degrees |
| 7 | GPS Height | km |
| 8 | GPS Vertical velocity | km/hr |
| 9 | Sample period | seconds |
| 10 | Steering angle | degrees |
| 11 | Wheel speed front left | rad/sec |
| 12 | Wheel speed front right | rad/sec |
| 13 | Wheel speed rear left | rad/sec |
| 14 | Wheel speed rear right | rad/sec |
| 15 | Yaw rate | deg/sec |
| 16 | Indicated vehicle speed | km/hr |
| 17 | Indicated longitudinal acceleration | g |
| 18 | Indicated lateral acceleration | g |
| 19 | Handbrake | activated or not (0 or 1) |
| 20 | Gear requested | number of gear employed (1-5) |
| 21 | Gear | number of gear employed (1-5) |
| 22 | Engine speed | rev/min |
| 23 | Coolant temperature | degree Celcius |
| 24 | Clutch position | activated or not (0 or 1) |
| 25 | Brake pressure | psi |
| 26 | Brake position | activated or not (0 or 1) |
| 27 | Battery voltage | volts |
| 28 | Air temperature | degrees Celcius |
| 29 | Accelerator pedal position | % activation |

Table 4 Environmental and driving scenarios investigated.

| No | Scenarios |
|---|---|
| 1 | Hard-brake |
| 2 | Sharp turn left and right |
| 3 | Swift maneuvers |
| 4 | Roundabout |
| 5 | Rain |
| 6 | Night and day |
| 7 | Skid |
| 8 | Mountain/hills |
| 9 | Dirt roads/ Gravel Roads |
| 10 | Country roads |
| 11 | Motorway |
| 12 | Town-centre driving |
| 13 | Traffic congestion |
| 14 | Successive left and right turns |
| 15 | Varying accelerations within a short duration |
| 16 | A -roads |
| 17 | B- roads |
| 18 | Wet roads |
| 19 | U-turns / Reverse drives |
| 20 | Mud road |
| 21 | Varying tyre pressure |
| 22 | Drifts |
| 23 | Bumps |
| 24 | Inner city driving |
| 25 | Winding roads |
| 26 | Zig-Zag drives |
| 27 | Approximate straight-line motion |
| 28 | Parking |
| 29 | Potholes |
| 30 | Residential roads |
| 31 | Stationary (No Motion) |
| 32 | Valleys |

Table 5 Information recorded from the smartphone sensors.

| No | Column Heading | Unit |
|---|---|---|
| 1 | GPS latitude | degrees |
| 2 | GPS longitude | degrees |
| 3 | GPS altitude | m |
| 4 | GPS speed | km/hr |
| 5 | GPS accuracy | m |
| 6 | GPS orientation | degrees |
| 7 | GPS satellites In range | N/A |
| 8 | Time since start | ms |
| 9 | Date | YYYY-MO-DD HH-MI-SS_SSS |
| 10 | Accelerometer X | m/s² |
| 11 | Accelerometer Y | m/s² |
| 12 | Accelerometer Z | m/s² |
| 13 | Gravity X | m/s² |
| 14 | Gravity Y | m/s² |
| 15 | Gravity Z | m/s² |
| 16 | Gyroscope (Yaw) | rad/s |
| 17 | Gyroscope (Pitch) | rad/s |
| 18 | Gyroscope (Roll) | rad/s |
| 19 | Magnetic field X | µT |
| 20 | Magnetic field Y | µT |
| 21 | Magnetic field Z | µT |
| 22 | Orientation (Yaw) | degrees |
| 23 | Orientation (Pitch) | degrees |
| 24 | Orientation (Roll) | degrees |

*Experiment Setup*

**Vehicle Experiment Setup**

The vehicle used for the data collection exercise was a front wheel drive Ford Fiesta Titanium as shown in Figure 2. A Racelogic VBOX Video HD2 was used to record the data from the vehicle CAN bus as well as the corresponding GPS coordinates at each sampling instance. As shown in Figures 1 and 2, the GPS antenna was placed centrally at the top of the vehicle to ensure optimal signal reception. The Racelogic VBOX Video HD2 CAN – Bus data logger (10 Hz) was used to record the data shown in Table 3 directly from the CAN bus of the vehicle with a sampling and update frequency of 10Hz.

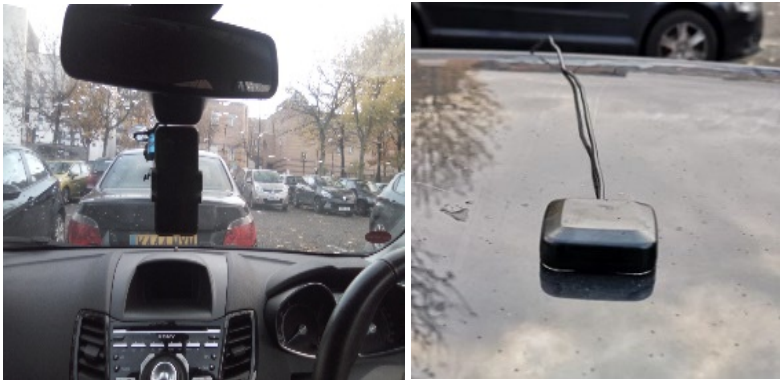

Figure 1 Smartphone and GPS antenna setup.

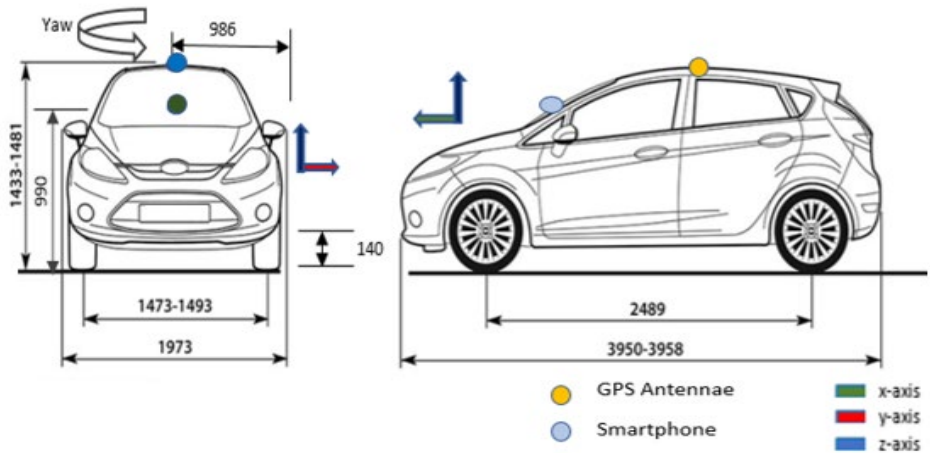

*Figure 2 Sensor locations and dimension of vehicle* [4]*.*

**Smartphone Measurement Setup**

A Ford Fiesta Titanium, Volvo XC70, Renault Mégane and Toyota Corolla Verso were used to collect the smartphone datasets. The smartphone was held with a phone holder attached to the vehicle as shown in Figure 1. Using the Androsensor app, all data were sampled every 0.1s with a GPS (smartphone) update rate of 1Hz. Figures 1 and 2 show the axis alignment of the smartphone sensors. The smartphone sensors employed were a 3-axis accelerometer, a 3-axis gyroscope, a 3-axis magnetometer and heading, as well as the GPS latitude and longitude coordinates all present within the phone. Other information such as the vehicle's velocity and acceleration were recorded from the smartphone's GPS. Table 5 highlights the data recorded from the smartphone data. The datasets described in Tables A1-1 to A5-2 were collected using the Huawei P20 pro smartphone.

*Acknowledgements*

We would like to thank Mr. Andy Thompson, Mr. Thierry Touzet, Miss. Sarah Tompkins, Mr. Yannick Weber, Dr. Maciej Cieslak, Mr. Felix Batsch and Google LLC for their help on this project.

*Declaration of Competing Interest*

The authors declare that they have no known competing financial interests or personal relationships which have, or could be perceived to have, influenced the work reported in this article.

*Ethics Statement*

The study and data collection have been approved by Coventry University Ethics Board under Project ID P95615.

*CRediT author statement*

**Uche Onyekpe:** Conceptualization, Methodology, Investigation, Validation, Writing - Original Draft, Writing - Review & Editing, Supervision; **Vasile Palade:** Investigation, Writing - Review & Editing; **Stratis Kanarachos:** Conceptualization, Investigation, Resources, Writing - Review & Editing; **Alicja Szkolnik:** Data Curation, Writing - Review & Editing.

*Appendix*

*Table A1-1 Dataset description from Driver A, B and C.*

| Driver | Dataset name | Features | Cities and towns covered | Weather conditions | Collection date | Velocity and acceleration range | Total time driven and distance covered | Total number of data points | Corresponding smartphone dataset |
|---|---|---|---|---|---|---|---|---|---|
| A | V-S1 | B-road (B4101), roundabout (x9), reverse (x5), hilly road, A4053 (ring-road), hard-brake, tyre pressure E | Coventry | 15 / 4 °C, Sunny, Humidity:73%, Wind:2.486 mph N | 08/09/2019 | 0.0 to 93.8 km/hr, -0.59 to 0.34 g | 86.3 mins, 38.16 km | 51790 | S-S1 |
| | V-S2 | B-road (B4112, B4065), roundabout (x18), reverse drive (x8), motorway, dirt road, u-turn (x5), country road, successive left-right turns, hard-brake, A-roads (A4600), tyre pressure E | Coventry, Nuneaton | 17 / 15 °C Passing clouds. Humidity:47% Wind:3.728 mph N | 08/09/2019 | 0.0 to 105.2 km/hr, -0.56 to 0.43 g | 156.5 mins, 75.64 km | 93900 | S-S2 |
| | V-S3a | Round-about (x15), u-turn/reverse drive (x4), motorway (M6), A-road (A4600, A426), hard-brake, swift maneuvers, country roads, change in speed, night-time, sharp turn left/right, tyre pressure E | Coventry, Rugby | 17 / 12 °C, Passing clouds. Humidity:65% Wind:6.836 mph W | 04/09/2019 | 0.0 - 98.0 km/hr, -0.57 to 0.4 g | 41.1 mins, 26.0 km | 24660 | S-S3a |
| | V-S3b | Successive left-right turns (x21), reverse/u-turns (x1), tyre pressure – E | Rugby | | 04/09/2019 | 0.0 to 44.8 km/hr, -0.37 to 0.3 g | 11.4 mins, 3.8 km | 6840 | S-S3b |
| | V-S3c | Roundabout(x4), A-road (A428), country roads, tyre pressure E | Rugby, Coventry | | 04/09/2019 | 0.0 to 117.1 km/hr, -0.36 to 0.35 g | 62.0 mins, 44.28 km | 37220 | S-S3c |
| | V-S4 | Roundabout (x14), u-turn, A-road, successive left-right turns, swift maneuvers, change in speed, night-time, A-road (A429, A45, A46), ring-road (A4053), tyre pressure E | Coventry | 13 / 12 °C, Passing clouds. Humidity:83% Wind:8.078 mph WNW | 06/09/2019 | 0.0 to 109.6 km/hr, -0.48 to 0.41 g | 163.0 mins, 93.9 km | 97824 | S-S4 |
| B | V-M | Roundabout (x30), successive left-right turns, hard-brake (x21), swift maneuvers (x5), country roads, sharp turn left/right, daytime, u-turn (x1), u-turn reverse (x7), tyre pressure E | Coventry | 15 / 12 °C, Partly sunny. Humidity:80% Wind:8.078 mph NW | 07/09/2019 | 0.0 to 100.7 km/hr, -1.01 to 0.44 g | 176.7 mins, 105.44 km | 105995 | S-M |
| C | V-St1 | Roundabout (x9), A-road (A452), B-road, car park navigation, tyre pressure E | Coventry, Kenilworth | 13 / 10 °C, Passing clouds. Humidity:56% Wind:7.457 mph ESE | 01/04/2019 | 0.0 to 73.3 km/hr, -0.39 to 0.45 g | 95.4 mins, 47.05 km | 57213 | N/A |
| Driver | Dataset name | Features | Cities and towns covered | Weather conditions | Collection date | Velocity and acceleration range | Total time driven and distance covered | Total number of data points | Corresponding smartphone dataset |

*Table A1-2 Dataset description from Driver C and D.*

| | | | | | | | | | |
|---|---|---|---|---|---|---|---|---|---|
| C | V-St4 | Roundabout (x1), A-road (A4114, A444, A46), motorway (M40), tyre pressure E | Coventry, Warwick, Chesterton | 9 / 4 °C Scattered clouds. Humidity:72% Barometer:991 mbar Wind:12.428 mph W | 04/03/2019 | 0.0 to 101.4 km/hr, -0.27 to 0.13 g | 22.7 mins, 28.48 km | 13591 | N/A |
| | V-St6 | Motorway (M40), daytime, tyre pressure E | Stokenchurch, Headington Oxford | 11 / 9 °C, Passing clouds. Humidity:62% Wind:10.564 mph SSW | 05/03/2019 | 0.0 to 122.1 km/hr, -0.32 to 0.35 g | 85.6 mins, 113.63 km | 51360 | N/A |
| | V-St7 | Motorway (M40), residential roads, A-road (A46), tyre pressure E | Stokenchurch, Headington Oxford, Coventry, Kenilworth, Warwick | 7 / 6 °C Light rain. Partly sunny. Humidity:85% Wind:14.914 mph W | 07/03/2019 | 0.0 to 117.9 km/hr, -0.3 to 0.3 g | 74.0 mins, 90.06 km | 44427 | N/A |
| D | V-Y1 | Roundabout (x20), successive left-right turns, hard-brake, swift maneuvers, sharp turn left/right, reverse/u-turn (x8), tyre pressure E | Coventry | 22 / 16 °C, Passing clouds. Humidity:74% Wind:6.836 mph SSW | 30/08/2019 | 0.0 to 87.5 km/hr, -0.85 to 0.36 g | 117.2 mins, 60.86 km | 70341 | S-Y1 |
| | V-Y2 | Roundabout (x9), u-turn/reverse (x1), A-road, B-road, country road, tyre pressure E | Coventry, Keniltworth | 7 / 6 °C Light rain. Partly sunny. Humidity:85% Wind:14.914 mph W | 08/03/2019 | 0.0 to 73.3 km/hr, -0.39 to 0.45 g | 95.4 mins, 47.05 km | 57213 | N/A |

*Table A2-1 Description of datasets V-Vta1a to V-Vta17 from Driver E.*

| Driver | Dataset name | Features | Cities and towns covered | Weather conditions | Collection date | Velocity and acceleration range | Total time driven and distance covered | Total number of data points | Corresponding smartphone dataset |
|---|---|---|---|---|---|---|---|---|---|
| E | V-Vta1a | Wet road, gravel road, country road, sloppy roads, roundabout (x3), hard-brake on wet road, tyre pressure A | Nuneaton, Walton on Trent | 4-10 / 3-6 °C Passing clouds, Broken Clouds, Scattered Clouds. Humidity:75-93% Wind:4.971 mph SE | 14/112019 | 0.0 to 103.4 km/hr, -0.54 to 0.35 g | 43.0 mins, 40.74 km | 25821 | S-Vta1a |
| | V-Vta1b | Hard-brake on muddy road, wet road, country road, tyre pressure A | Coton in the Elms, Walton on Trent | | | 0.1 to 77.7 km/hr, -0.49 to 0.28 g | 1.6 mins, 1.26 km | 956 | S-Vta1b |
| | V-Vta2 | Roundabout (x2), A-road (A511, A5121, A444), country road, hard-brakes, tyre pressure A | Walton on Trent, Burton on Trent | | | 0.0 to 81.6 km/hr, -0.59 to 0.38 g | 18.3 mins, 11.07 km | 10995 | S-Vta2 |
| | V-Vta3 | Roundabout (x1), swift maneuvers, tyre pressure A | Burton on Trent | | | 0.0 to 45.8 km/hr, -0.31 to 0.27 g | 1.5 mins, 0.38 km | 875 | S-Vta3 |
| | V-Vta4 | A-road (A511), tyre pressure A | Burton on Trent | | | 5.9 to 51.7 km/hr, -0.37 to 0.28 g | 3.0 mins, 2.02 km | 1809 | S-Vta4 |
| | V-Vta5 | Roundabout (x1), A-road (A511), tyre pressure A | Burton on Trent | | | 29.2 to 51.1 km/hr, -0.26 to 0.09 g | 0.6 min, 0.42 km | 357 | S-Vta5 |
| | V-Vta6 | A-road (A511), tyre pressure A | Burton on Trent | | | 43.8 to 103.9 km/hr, -0.24 to 0.13 g | 2.3 mins, 2.62 km | 1393 | S-Vta6 |
| | V-Vta7 | Roundabout (x2), A-road (A511), hard-brakes, tyre pressure A | Burton on Trent | | | 22.4 to 113.1 km/hr, -0.54 to 0.18 g | 1.4 mins, 1.54 km | 857 | S-Vta7 |
| | V-Vta8 | Town roads, A-roads (A511), tyre pressure A | Hatton Derby | | | 0.0 to 77.6 km/hr, -0.45 to 0.3 g | 6.2 mins, 3.43 km | 3697 | S-Vta8 |
| | V-Vta9 | Hard-brakes, A–road (A50), tyre pressure A | Derby | | | 48.9 to 87.7 km/hr, -0.6 to 0.14 g | 0.4 min, 0.43 km | 226 | S-Vta9 |
| | V-Vta10 | Roundabout (x1), A-road (A50), tyre pressure A | Sudbury Ashburne | | | 38.8 to 118.0 km/hr, -0.28 to 0.13 g | 2.6 mins, 3.95 km | 1570 | S-Vta10 |
| | V-Vta11 | Roundabout (x2), A-road (A50), tyre pressure A | Oaks Green Ashburne | | | 26.8 to 97.7 km/hr, -0.45 to 0.15 g | 1.0 min, 0.92 km | 589 | S-Vta11 |
| | V-Vta12 | changes in acceleration in a short period of time, A-road (A515), tyre pressure A | Ashburne | | | 44.7 to 85.3 km/hr, -0.44 to 0.13 g | 1.1 mins, 1.27 km | 690 | S-Vta12 |
| | V-Vta13 | A-road (A515), country road, hard-brakes, tyre pressure A | Ashburne | | | 72.7 to 103.6 km/hr, -0.38 to 0.12 g | 0.8 mins, 1.14 km | 473 | S-Vta13 |
| | V-Vta14 | Hard-brakes, changes in acceleration in a short period of time, A-road (A515), tyre pressure A | Ashburne | | | 52.8 to 91.0 km/hr, -0.32 to 0.13 g | 4.8 mins, 5.45 km | 2893 | S-Vta14 |
| | V-Vta15 | A–road (A515), tyre pressure A | Ashburne | | | 60.1 to 78.8 km/hr, -0.12 to 0.06 g | 1.4 mins, 1.72 km | 869 | S-Vta15 |
| | V-Vta16 | Roundabout (x3), hilly roads, country road, A-road (A515), tyre pressure A | Thorpe Ashburne | | | 0.0 to 93.9 km/hr, -0.49 to 0.42 g | 18.9 mins, 13.72 km | 11361 | S-Vta16 |
| | V-Vta17 | Hilly roads, hard-brake, stationary (no motion), tyre pressure A | Ilam, Blore | | | 0.0 to 56.2 km/hr, -0.51 to 0.28 g | 7.7 mins, 4.19 km | 4594 | S-Vta17 |

*Table A2-2 Description of datasets V-Vta19 to V-Vta30 from Driver E.*

| Driver | Dataset name | Features | Cities and towns covered | Weather conditions | Collection date | Velocity and acceleration range | Total time driven and distance covered | Total number of data points | Corresponding smartphone dataset |
|---|---|---|---|---|---|---|---|---|---|
| E | V-Vta19 | Hilly road, tyre pressure A | Ilam | 4-10 / 3-6 °C Passing clouds, Broken Clouds, Scattered Clouds. Humidity:75-93% SE Wind:4.971 mph | 06/112019 | 0.0 to 55.2 km/hr, -0.35 to 0.22 g | 0.5 min, 0.26 km | 310 | S-Vta19 |
| | V-Vta20 | Hilly road, approximate straight-line travel, tyre pressure A | Ilam | | | 0.0 to 44.8 km/hr, -0.19 to 0.3 g | 5.4 mins, 0.39 km | 3223 | S-Vta20 |
| | V-Vta21 | Hilly road, tyre pressure A | Ilam | | | 0.0 to 74.8 km/hr, -0.44 to 0.24 g | 3.5 mins, 2.76 km | 2088 | S-Vta21 |
| | V-Vta22 | Hilly road, hard-brake, tyre pressure A | Ilam | | | 14.8 to 55.8 km/hr, -0.53 to 0.16 g | 2.6 mins, 1.67 km | 1572 | S-Vta22 |
| | V-Vta23 | Hilly road, hard-brake, tyre pressure A | Thorpe | | | 0.0 to 51.9 km/hr, -0.57 to 0.42 g | 1.9 mins, 1.1 km | 1119 | S-Vta23 |
| | V-Vta24 | Hilly road, tyre pressure A | Thorpe | | | 0.0 to 56.4 km/hr, -0.46 to 0.36 g | 2.0 mins, 0.71 km | 1184 | S-Vta24 |
| | V-Vta25 | U-turn, tyre pressure A | Thorpe | | | 0.0 to 48.6 km/hr, -0.46 to 0.3 g | 1.1 mins, 0.16 km | 646 | S-Vta25 |
| | V-Vta26 | Gravel road, dirt road, hilly road, tyre pressure A | Thorpe | | | 0.0 to 55.1 km/hr, -0.27 to 0.44 g | 3.2 mins, 1.02 km | 1947 | S-Vta26 |
| | V-Vta27 | Gravel road, several hilly roads, potholes, country road, A-road (A515), tyre pressure A | Ashburne | | | 0.0 to 65.0 km/hr, -0.43 to 0.29 g | 4.8 mins, 3.16 km | 2853 | S-Vta27 |
| | V-Vta28 | Country road, hard-brakes, valley, A-road (A515), tyre pressure A | Milldale | | | 0.0 to 66.0 km/hr, -0.58 to 0.31 g | 7.0 mins, 3.94 km | 4219 | S-Vta28 |
| | V-Vta29 | Hard-brakes, country road, hilly road, windy road, dirt road, wet road, reverse drive (x2), bumps, rain, B-road (B5053), country road, u-turn (x3), windy road, valley, tyre pressure A | Wetton, Milldale | | | 0.0 to 102.0 km/hr, -0.8 -to 0.38 g | 39.6 mins, 26.12 km | 23737 | S-Vta29 |
| | V-Vta30 | Rain, wet road, u-turn (x2), A-road (A53, A515), inner town driving, B-road (B5053), tyre pressure A | Buxton | | | 0.0 to 100.0 km/hr, -0.47 to 0.36 g | 28.6 mins, 11.77 km | 17179 | S-Vta30 |

*Table A3 Description of datasets V-Vtb1 to V-Vtb13 from Driver E.*

| Driver | Dataset name | Features | Cities and towns covered | Weather conditions | Collection date | Velocity and acceleration range | Total time driven and distance covered | Total number of data points | Corresponding smartphone dataset |
|---|---|---|---|---|---|---|---|---|---|
| E | V-Vtb1 | Valley, rain, wet road, country road, u-turn (x2), hard-brake, swift maneuver, A–road (A6, A6020, A623, A515), B-road (B6405), round about (x3), daytime, tyre pressure A | Bakewell, Tideswell, Ashford on water, Buxton | 4-8 / 4 °C Rain, Passing clouds, Broken Clouds, Chilly. Humidity:94-98% Barometer:1004 mbar N Wind:10.564 mph | 06/11/2019 | 0.0 to 101.2 km/hr, -0.63 to 0.36 g | 54.1 mins, 41.94 km | 32459 | S-Vtb1 |
|  | V-Vtb2 | Country road, wet road, dirt road, tyre pressure A | Youlgreave |  |  | 0.0 to 61.1 km/hr, -0.36 to 0.39 g | 9.5 mins, 4.35 km | 5712 | S-Vtb2 |
|  | V-Vtb3 | Reverse, wet road, dirt road, gravel road, night-time, tyre pressure A | Youlgreave |  |  | 0.0 to 37.5 km/hr, -0.23 to 0.33 g | 13.8 mins, 0.71 km | 8289 | S-Vtb3 |
|  | V-Vtb4 | Dirt road, country road, gravel, wet road, tyre pressure A | Youlgreave |  |  | 0.0 to 32.7 km/hr, -0.31 to 0.27 g | 1.0 min, 0.27 km | 625 | S-Vtb4 |
|  | V-Vtb5 | Dirt road, country road, gravel road, hard-brakes, Wet road, B-road (B6405, B6012, B5056), inner-town driving, A-road, motorway (M42, M1), rush hour(traffic), round-about (x6), a-road (A5, A42, A38, A615, A6), tyre pressure A | Atherstone, Nuthall, Hilcote, Matlock, Rowsley, Youlgreave |  |  | 0.0 to 112.9 km/hr, -0.55 to 0.42 g | 107.7 mins, 111.66 km | 64610 | S-Vtb5 |
|  | V-Vtb6 | A-road (A5), tyre pressure A | Atherstone |  |  | 52.7 to 73.0 km/hr, -0.11 to 0.11 g | 0.8 min, 0.89 km | 508 | S-Vtb6 |
|  | V-Vtb7 | Approximate straight-line motion, night-time, A-road (A5), tyre pressure A | Atherstone |  |  | 29.1 to 69.2 km/hr, -0.37 to 0.13 g | 0.8 min, 0.72 km | 461 | S-Vtb7 |
|  | V-Vtb8 | Approximate straight-line motion, nighttime, wet road, A-road (A5), tyre pressure A | Atherstone |  |  | 60.9 to 76.5 km/hr, -0.35 to 0.08 g | 1.2 mins, 1.35 km | 699 | S-Vtb8 |
|  | V-Vtb9 | Approximate straight-line motion, night-time, wet road, hard-brakes, A-road (A5), tyre pressure A | Nuneaton |  |  | 66.8 to 92.0 km/hr, -0.14 to 0.1 g | 0.8 min, 0.98 km | 457 | S-Vtb9 |
|  | V-Vtb10 | Round-about, wet road, night-time, A-road (A5), tyre pressure A | Nuneaton |  |  | 26.1 to 58.5 km/hr, -0.24 to 0.12 g | 0.3 min, 0.23 km | 195 | S-Vtb10 |
|  | V-Vtb11 | Approximate straight-line motion, night-time, wet road, A-road (A5), tyre pressure A | Nuneaton |  |  | 65.1 to 75.3 km/hr, -0.05 to 0.12 g | 0.7 min, 0.84 km | 433 | S-Vtb11 |
|  | V-Vtb12 | Roundabout (x1), wet road, night-time, tyre pressure A | Nuneaton |  |  | 22.2 to 71.6 km/hr, -0.38 to 0.17 g | 0.8 min, 0.61 km | 490 | S-Vtb12 |
|  | V-Vtb13 | Parking, wet road, tyre pressure A | Nuneaton |  |  | 7.5 to 43.3 km/hr, -0.31 to 0.22 g | 2.1 mins, 0.99 km | 1245 | N/A |

*Table A4 Description of datasets V-Vw1 to V-Vw12 from Driver E.*

| Driver | Dataset name | Features | Cities and towns covered | Weather conditions | Collection date | Velocity and acceleration range | Total time driven and distance covered | Total number of data points | Corresponding smartphone dataset |
|---|---|---|---|---|---|---|---|---|---|
| E | V-Vw1 | Stationary (no motion, sensor bias estimation), daytime, tyre pressure C | Nuneaton | 10 °C Smoke. Wind: 6 mph N Humidity: 86% | 08/01/2020 | 0.00 to 0.00 km/hr, 0.00 to -0.00 g | 34.1 mins, 0.00 km | 20475 | S-Vw1 |
| | V-Vw2 | A-road (A5, A421), motorway (M5), daytime, roundabout (x22), u-turn (x2), inner city driving, tyre pressure C | Nuneaton, Hinckley Milton Keynes | | | 0.0 to 115.4 km/hr, -0.62 to 0.45 g | 87.9 mins, 98.63 km | 52712 | S-Vw2 |
| | V-Vw3 | Roundabout (x6), daytime, B-road, inner-city driving, tyre pressure C | Milton Keynes | | | 0.0 to 77.4 km/hr, -0.47 to 0.41 g | 6.6 mins, 5.05 km | 3942 | S-Vw3 |
| | V-Vw4 | Roundabout (x77), swift-maneuvers, hard-brake, inner city driving, reverse, A-road, motorway (M5, M40, M42), country road, successive left-right turns, daytime, u-turn (x3), tyre pressure D | Milton Keynes, Buckingham, Droitwich Spa, Kidderminster, Worcester | | | 0.0 to 131.9 km/hr, -0.66 to 0.45 g | 211.0 mins, 214.62 km | 126573 | S-Vw4 |
| | V-Vw5 | Successive left-right turns, daytime, sharp turn left/right, tyre pressure D | Worcester | 10 °C Passing clouds. Wind: 2 mph N Humidity: 88% | | 0.0 to 38.7 km/hr, -0.4 to 0.21 g | 1.8 mins, 0.7 km | 1050 | S-Vw5 |
| | V-Vw6 | Bumps, swift-maneuvers, daytime, sharp turn left/right, pressure D | Worcester | | | 3.3 to 40.7 km/hr, -0.34 to 0.26 g | 2.1 mins, 1.08 km | 1288 | S-Vw6 |
| | V-Vw7 | Successive left-right turns, daytime, sharp turn left/right, tyre pressure D | Worcester | | | 0.4 to 42.2 km/hr, -0.37 to 0.37 g | 2.8 mins, 1.23 km | 1689 | S-Vw7 |
| | V-Vw8 | Successive left-right turns, daytime, sharp turn left/right, tyre pressure D | Worcester | | | 0.0 to 46.4 km/hr, -0.37 to 0.27 g | 2.7 mins, 1.12 km | 1599 | S-Vw8 |
| | V-Vw9 | Swift-maneuvers, daytime, hard-brake, tyre pressure D | Worcester | | | 3.8 to 42.0 km/hr, -0.67 to 0.21 g | 1.0 min, 0.45 km | 601 | S-Vw9 |
| | V-Vw10 | Hilly road, daytime, pressure D | Worcester | | | 11.8 to 58.9 km/hr, -0.42 to 0.11 g | 1.1 mins, 0.74 km | 670 | S-Vw10 |
| | V-Vw11 | Motorway (M5), daytime, roundabout (x5), tyre pressure D | | | | 0.0 to 98.4 km/hr, -0.37 to 0.33 g | 8.2 mins, 5.85 km | 4924 | S-Vw11 |
| | V-Vw12 | Approximate straight-line motion, daytime, Motorway (M5), tyre pressure D | | 7 °C Drizzle. Fog. Wind: 5 mph N Humidity: 93% | | 82.6 to 97.4 km/hr, -0.06 to 0.07 g | 1.75 mins, 2.64 km | 1050 | S-Vw12 |

*Table A5 Description of datasets V-Vw13 to V -Vw17 from Driver E.*

| Driver | Dataset name | Features | Cities and towns covered | Weather conditions | Collection date | Velocity and acceleration range | Total time driven and distance covered | Total number of data points | Corresponding smartphone dataset |
|---|---|---|---|---|---|---|---|---|---|
| E | V-Vw13 | Approximate straight-line motion, daytime, motorway (M5), tyre pressure D | | 7 °C Drizzle. Fog. Wind: 5 mph N Humidity: 93% | 08/01/2020 | 94.0 to 115.0 km/hr, -0.07 to 0.06 g | 0.5 min, 0.82 km | 297 | S-Vw13 |
| | V -Vw14a | Motorway (M5), nighttime, tyre pressure D | | | | 61.9 to 109.4 km/hr, -0.38 to 0.12 g | 5.2 mins, 7.92 km | 3140 | S-Vw14a |
| | V -Vw14b | Motorway (M42), nighttime, tyre pressure D | | | | 12.6 to 120.1 km/hr, -0.28 to 0.28 g | 32.7 mins, 41.21 km | 19600 | S-Vw14b |
| | V -Vw14c | Motorway (M42), roundabout (x2), A-road (A446), nighttime, hard-brakes, tyre pressure D | | | | 0.0 to 100.5 km/hr, -0.53 to 0.41 g | 26.4 mins, 17.15 km | 15857 | S-Vw14c |
| | V -Vw15 | Stationary (no motion, sensor bias estimation), nighttime, tyre pressure D | Dordon | 8 °C Cool. Wind: 2 mph N Humidity: 80% | | 0.0 to 0.0 km/hr, 0.00 to 0.0 g | 2.3 mins, 0.00 km | 1391 | S-Vw15 |
| | V -Vw16a | A–road (A5), roundabout (x2), tyre pressure D | Atherstone | 8 °C Rain showers. Overcast. 2 mph N 80% | | 0.0 to 83.5 km/hr, -0.39 to 0.4 g | 10.0 mins, 8.49 km | 6000 | S-Vw16a |
| | V -Vw16b | Hard-brakes, nighttime, A-road (A5), approximate straight-line travel, tyre pressure D | Nuneaton | | | 1.3 to 86.3 km/hr, -0.75 to 0.29 g | 2.0 mins, 1.99 km | 1171 | S-Vw16b |
| | V -Vw17 | Hard-brakes, nighttime, A-road (A5), approximate straight-line travel, tyre pressure D | Calcedote | | | 31.5 to 72.7 km/hr, -0.8 to 0.19 g | 0.5 min, 0.54 km | 329 | S-Vw17 |

*Table A6-1 Description of datasets V-Vfa01to V-Vfb02c from Driver E.*

| Driver | Dataset name | Features | Cities and towns covered | Weather conditions | Collection date | Velocity and acceleration range | Total time driven and distance covered | Total number of data points | Corresponding smartphone dataset |
|---|---|---|---|---|---|---|---|---|---|
| E | V-Vfa01 | A-road (A444), roundabout (x1), B–road (B4116), daytime, hard-brakes, tyre pressure A | Nuneaton, Twycross, Measham | 6 °C Quite cool. Wind: 8 mph N Humidity: 97% | 08/11/2019 | 0.0 to 98.4 km/hr, -0.56 to 0.42 g | 19.2 mins, 18.8 km | 11535 | S-Vfa01 |
| | V-Vfa02 | B-road (B4116), roundabout (x5), A-road (A42, A641), motorway (M1, M62), high rise buildings, hard-brake, tyre pressure C | Bradford, Measham | 7 °C, Scattered clouds. Wind: 8 mph N Humidity: 87% | | 0.0 to 117.9 km/hr, -0.67 to 0.48 g | 112.9 mins, 163.38 km | 67755 | S-Vfa02 |
| | V-Vfb01a | City-centre driving, roundabout (x1), wet road, ring-road, nighttime, tyre pressure C | Bradford | 5 °C, Light rain. Passing clouds. Wind: 10 mph N Humidity:87% | | 0.0 to 68.9 km/hr, -0.43 to 0.42 g | 28.3 mins, 6.81 km | 17000 | N/A |
| | V-Vfb01b | Motorway (M606), round-about (x1), city roads, traffic, wet road, changes in acceleration in short periods of time, nighttime, tyre pressure C | | | | 0.0 to 83.0 km/hr, -0.38 to 0.23 g | 6.5 mins, 4.07 km | 3880 | N/A |
| | V-Vfb01c | Motorway (M62), wet-road, heavy traffic, nighttime, tyre pressure C | | | | 0.2 to 104.5 km/hr, -0.36 to 0.38 g | 10.5 mins, 10.66 km | 6320 | N/A |
| | V-Vfb01d | Roundabout (x1), A-road (A650), nighttime, tyre pressure C | | | | 0.0 to 56.0 km/hr, -0.46 to 0.36 g | 17.9 mins, 3.39 km | 10713 | N/A |
| | V-Vfb02a | Motorway (M1), roundabout (x2), A-road (A650), nighttime, hard-brakes, tyre pressure D | East Ardsley, | 7 °C, Rain showers. Overcast. Wind: 12 mph N Humidity:86% | | 0.0 to 122.3 km/hr, -0.5 to 0.37 g | 59.9 mins, 96.5 km | 35960 | N/A |
| | V-Vfb02b | Roundabout (x1), bumps, successive left-right turns, hard-brakes (x7), swift-maneuvers, nighttime, tyre pressure D | Nuthall | | | 0.0 to 84.3 km/hr, -0.5 to 0.35 g | 18.3 mins, 7.69 km | 11000 | N/A |
| | V-Vfb02c | U-turn (x1), hard-brakes, nighttime, tyre pressure D | Nuthall | | | 2.0 to 52.8 km/hr, -0.53 to 0.26 g | 1.1 mins, 0.54 km | 640 | N/A |

*Table A6-2 Description of datasets V-Vfb02d to V-Vfb02g from Driver E.*

| Driver | Dataset name | Features | Cities and towns covered | Weather conditions | Collection date | Velocity and acceleration range | Total time driven and distance covered | Total number of data points | Corresponding smartphone dataset |
|---|---|---|---|---|---|---|---|---|---|
| E | V-Vfb02d | Round-about (x1), nighttime, tyre pressure D | Nuthall | 7 °C, Rain showers. Overcast. Wind: 12 mph N Humidity:86% | 08/11/2019 | 0.0 to 57.3 km/hr, -0.33 to 0.31 g | 1.5 mins, 0.84 km | 880 | N/A |
| | V-Vfb02e | Changes in acceleration in short period of time, nighttime, tyre pressure D | Nuthall | | | 37.4 to 73.9 km/hr, -0.24 to 0.19 g | 1.6 mins, 1.52 km | 980 | N/A |
| | V-Vfb02f | Roundabout (x1), nighttime, tyre pressure D | Nuthall | | | 1.6 to 49.5 km/hr, -0.24 to 0.32 g | 1.1 mins, 0.47 km | 660 | N/A |
| | V-Vfb02g | Motorway (M1), A-road (A42, A444, A5), country road, roundabout (x2), hard-brakes, nighttime, tyre pressure D | Nuneaton | | | 0.0 to 119.4 km/hr, -0.51 to 0.35 g | 45.3 mins, 63.56 km | 27159 | N/A |

*Table A7 Information on other Smartphone Dataset captured independently from drivers F, G and H.*

| Driver | Dataset name | Location | Comments | Vehicle model | Phone Model | Total Time driven (mins) | Total distance covered (km) | Total number of data points |
|---|---|---|---|---|---|---|---|---|
| F | S-T1, S-T2, S-T3, S-T4, S-T5, S-T6, S-T8, S-T9 | France | **Information on 3-axis orientation and magnetic field not available.** | Renault Megane | Motorola moto G7 power | 1005.70 | 1508.39 | 603425 |
| | S-T10, S-T11 | France | - | Renault Megane | Motorola moto G7 power | 20.60 | 8.86 | 12389 |
| G | S-I | Nigeria | - | Toyota Corolla Verso | Huawei P20 pro, | 9.70 | 0.06 | 5800 |
| H | S-A1, S-A2, S-A3, S-A4, S-A5, S-A6, S-A7, S-A8, S-A9, S-A10, S-A11, S-A12, S-A13 | England | - | Volvo XC70 | Blackberry Priv | 638.30 | 1511.93 | 382956 |